# A New All-Digital Background Calibration Technique for Time-Interleaved ADC Using First Order Approximation FIR Filters

Jiadong Hu, Zhe Cao, Qi An, Lei Zhao, Shubin Liu

**Abstract**—This paper describes a new all-digital technique for calibration of the mismatches in time-interleaved analog-to-digital converters (TIADCs) to reduce the circuit area. The proposed technique gives the first order approximation of the gain mismatches and sample-time mismatches, and employs first order approximation FIR filter banks to calibrate the sampled signal, which do not need large number of FIR taps. In the case of a two-channel 12-bit TIADC, the proposed technique improves SINAD of simulated data from 45dB to 69dB, and improves SINAD of measured data from 47dB to 53dB, while the number of FIR taps is only 30. In the case of slight mismatches, 24-bit FIR coefficient is sufficient to correct 12-bit signals, which makes it easy to implement this technique in hardware. In addition, this technique is not limited by the number of sub-ADC channels and can be calculated in parallel in hardware, these features enable this technique to be versatile and capable of real-time calibration.

## I. INTRODUCTION

IN many fields such as wireless infrastructure, wideband microwave backhauls and measurement equipment, high speed and high performance analog-to-digital converters (ADCs) play an essential role. To meet the requirements of speed and performance, time-interleaved analog-to-digital converter (TIADC) is proposed to be an effective architecture, by combining several slow but accurate sub-ADCs in parallel [1]. However, duo to discrepancies among the sub-ADCs, channel mismatches including offset, gain and sample-time mismatches distort the sampled signal and degrade the SINAD performance of the TIADC significantly [2]. Therefore, calibration methods are required to handle the channel mismatches and restore the dynamic performance of sub-ADCs.

Recently, a large number of all-digital calibration techniques of TIADC have been proposed. Some use adaptive blind calibration methods [3-5], some use an extra low-resolution ADC and a time-varying filter [6], some eliminate mismatches by means of signal subtraction between adjacent channels [7], and some use Hadamard transform and pseudo aliasing signal

[8-10]. This paper presents a new all-digital background calibration technique for the TIADC, consuming few hardware resources while compensating channel mismatches. The proposed technique uses FIR filters to complete the calibration function, but does not need many FIR taps. Thus, this method effectively reduces the amount of computation required for the calibration, while maintaining the dynamic performance of the TIADC. Moreover, this method is not limited by the number of TIADC channels and can be calculated in parallel in hardware.

## II. RESIDUAL ALIASING DUE TO GAIN AND SAMPLE-TIME MISMATCHES

Fig. 1 shows a block diagram of the M-channel TIADC. The sampling rate of the TIADC is $f_s=1/T_s$, and the sampling rate of each sub-ADC is $f_s$ /M. The digital outputs of sub-ADCs are combined to be the output of TIADC.

Fig. 2 shows a simplified block diagram of the two-channel TIADC. Each sub-ADC samples at a low frequency of $1/T_1$, while $T_1=2T_s$, and the corresponding discrete-time frequency is $\omega=\Omega T_1$ . $H_m(j\Omega)(m=0,1)$ is m-th channel frequency response with gain and sample-time mismatches. These channel responses are written as

$$H_0(j\Omega)=(1+\Delta g_0)e^{j\Omega(\Delta t_0 \cdot T_s)}$$
$$H_1(j\Omega)=(1+\Delta g_1)e^{j\Omega(T_s+\Delta t_1 \cdot T_s)} \quad (1)$$

or

$$H_0(j\frac{\omega}{2T_s})=(1+\Delta g_0)e^{j\varphi\frac{\Delta t_0}{M}}$$
$$H_1(j\frac{\omega}{2T_s})=(1+\Delta g_1)e^{j\varphi\frac{(1+\Delta t_1)}{M}} \quad (2)$$

where $\Delta g_m$ is the gain mismatch and $\Delta t_m$ is sample-time mismatch in m-th channel. Employ $\Delta o_m$ as offset mismatch, the discrete Fourier transform (DFT) of sampled signal $\hat{a}_m[n]$

This work was supported by the National Natural Science Foundation of China under Grant 11505182.

The authors are with the State Key Laboratory of Particle Detection and Electronics, University of Science and Technology of China; and Department of Modern Physics, University of Science and Technology of China, Hefei 230026, China (Corresponding author: Zhe Cao, e-mail: caozhe@ustc.edu.cn).

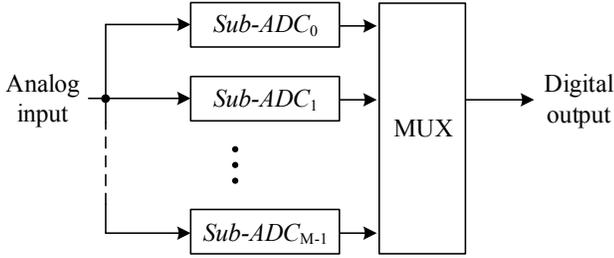

Fig. 1. M-channel TIADC block diagram.

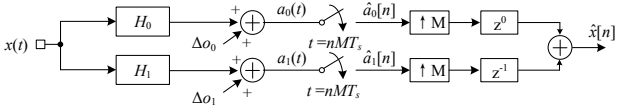

Fig. 2. A two-channel TIADC model.

are

$$\hat{A}_0(j\omega) = \frac{1}{2T_s} \sum_{k=-\infty}^{+\infty} X\left(j\frac{\omega}{2T_s} - j\frac{2\pi k}{2T_s}\right)$$
$$\cdot H_0\left(j\frac{\omega}{2T_s} - j\frac{2\pi k}{2T_s}\right)$$
$$+ \frac{2\pi\Delta o_0}{2T_s}\sum_{l=-\infty}^{+\infty}\delta(\omega - 2\pi l) \quad (3)$$

$$\hat{A}_1(j\omega) = \frac{1}{2T_s} \sum_{k=-\infty}^{+\infty} X\left(j\frac{\omega}{2T_s} - j\frac{2\pi k}{2T_s}\right)$$
$$\cdot H_1\left(j\frac{\omega}{2T_s} - j\frac{2\pi k}{2T_s}\right)$$
$$+ \frac{2\pi\Delta o_1}{2T_s}\sum_{l=-\infty}^{+\infty}\delta(\omega - 2\pi l)$$

where $X(j\frac{\omega}{2T_s})$ is the DFT of the analog input signal $x(t)$. The $\Delta g_m$, $\Delta t_m$ and $\Delta o_m$ can be calculated through a four-parameter sine wave fitting method [11].

Normally, offset mismatches can be simply calibrated by subtracting a certain constant, so we can assume that the offset mismatches $\Delta o_m = 0$ in formula derivation. For the two-channel TIADC, we assume that channel 0 is ideal, which means $\Delta g_0 = 0$, $\Delta t_0 = 0$. After ignoring offset mismatches and assuming channel 0 to be ideal, the DFT of sampled signal in (3) are simplified to

$$\hat{A}_0(j\omega) = \hat{A}_0^{ideal}(j\omega)$$
$$\hat{A}_1(j\omega) = \frac{1}{2T_s} \sum_{k=-\infty}^{+\infty} X\left(j\frac{\omega}{2T_s} - j\frac{2\pi k}{2T_s}\right) H_1\left(j\frac{\omega}{2T_s} - j\frac{2\pi k}{2T_s}\right). \quad (4)$$

The digital output of sub-ADC1 has aliasing signals because of the non-ideality of $H_1(j\Omega)$. Assuming that $\Delta g_1$ and $\Delta t_1$ are much less than 1 in an actual TIADC, by the first order approximation, $H_1(j\Omega)$ is simplified to

$$H_1\left(j\frac{\omega}{2T_s}\right) \approx H_1^{ideal}\left(j\frac{\omega}{2T_s}\right) \cdot \left(1 + \Delta g_1 + j\omega\frac{\Delta t_1}{2}\right) \quad (5)$$

and the DFT of sampled signal in (3) are simplified to

$$\hat{A}_0(j\omega) = \hat{A}_0^{ideal}(j\omega)$$
$$\hat{A}_1(j\omega) \approx \hat{A}_1^{ideal}(j\omega) \cdot \{1 + \Delta g_1 \quad (6)$$
$$+ \sum_{k=-\infty}^{+\infty} j(\omega - 2\pi k)\frac{\Delta t_1}{2}\Phi(\omega - 2\pi k)\}$$

where

$$\Phi(\omega) = \begin{cases} 1, & \omega \in (-\pi, \pi) \\ 0, & \omega \notin (-\pi, \pi) \end{cases}. \quad (7)$$

The second term and the third term in (6) show residual aliasing signals due to the gain mismatch and the sample-time mismatch, respectively.

## III. PROPOSED CALIBRATION TECHNIQUE

Fig. 3 shows the proposed calibration architecture in two-channel TIADC. The key idea is using the first order approximation FIR filter to reduce the channel mismatches. After calibration, the final output is approximately equal to the ideal output.

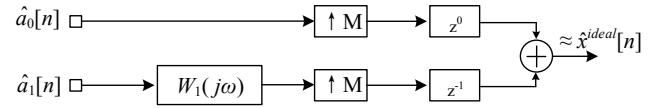

Fig. 3. The proposed calibration architecture in two-channel TIADC.

### A. Aliasing Elimination

The residual aliasing signal in channel 1 can be eliminated by means of the following manner

$$\hat{A}_1(j\omega) - \Delta g_1 \hat{A}_1(j\omega) - \hat{A}_1(j\omega) \cdot \sum_{k=-\infty}^{+\infty} j(\omega - 2\pi k)\frac{\Delta t_1}{2}\Phi(\omega - 2\pi k)$$
$$= \hat{A}_1^{ideal}(j\omega) \cdot \left(1 - \left(\Delta g_1 + \sum_{k=-\infty}^{+\infty} j(\omega - 2\pi k)\frac{\Delta t_1}{2}\Phi(\omega - 2\pi k)\right)^2\right) \quad (8)$$
$$\approx \hat{A}_1^{ideal}(j\omega)$$

Suppose first order approximation FIR filter

$$w_1[n] = \begin{cases} \dfrac{(-1)^{n+1}}{n} \cdot \dfrac{\Delta t_1}{2}, & n \neq 0, n \in Z \\ 1 - \Delta g_1, & n = 0 \end{cases} \quad (9)$$

$$\xrightarrow{DFT} W_1(j\omega) = 1 - \Delta g_1 - \sum_{k=-\infty}^{+\infty} j(\omega - 2\pi k)\frac{\Delta t_1}{2}\Phi(\omega - 2\pi k)$$

then

$$\hat{a}_1[n] * w_1[n] \approx \hat{a}_1^{ideal}[n]$$
$$\xleftarrow{DFT} \hat{A}_1(j\omega) \cdot W_1(j\omega) \approx \hat{A}_1^{ideal}(j\omega) \quad . \quad (11)$$

### B. Proposed Technique in M-Channel TIADC

For the M-channel TIADC, the DFT of sampled signal in (3) are rewritten as

$$\hat{A}_m(j\omega) \approx \hat{A}_m^{ideal}(j\omega) \cdot \{1 + \Delta g_m$$
$$+ \sum_{k=-\infty}^{+\infty} j(\omega - 2\pi k)\frac{\Delta t_m}{M}\Phi(\omega - 2\pi k)\} \quad (12)$$

where $m=0, 1, \dots, (M-1)$. The first order approximation FIR filters are rewritten as

$$w_m[n] = \begin{cases} \dfrac{(-1)^{n+1}}{n} \cdot \dfrac{\Delta t_m}{M}, & n \neq 0, n \in Z \\ 1 - \Delta g_m, & n = 0 \end{cases} \quad (13)$$

$$\overset{DFT}{\longleftrightarrow} W_m(j\omega) = 1 - \Delta g_m - \sum_{k=-\infty}^{+\infty} j(\omega - 2\pi k)\dfrac{\Delta t_m}{M} \cdot \Phi(\omega - 2\pi k)$$

Since $\Delta g_m$ is far less than 1, the first order approximation FIR filters can also be written as

$$w_m[n] = \begin{cases} \dfrac{(-1)^{n+1}}{n} \cdot \dfrac{\Delta t_m}{M}, & n \neq 0, n \in Z \\ \dfrac{1}{1 + \Delta g_m}, & n = 0 \end{cases} \quad (14)$$

$$\overset{DFT}{\longleftrightarrow} W_m(j\omega) = \dfrac{1}{1 + \Delta g_m} - \sum_{k=-\infty}^{+\infty} j(\omega - 2\pi k)\dfrac{\Delta t_m}{M} \cdot \Phi(\omega - 2\pi k)$$

then

$$\hat{a}_m[n] * w_m[n] \approx \hat{a}_m^{ideal}[n]$$
$$\overset{DFT}{\longleftrightarrow} \hat{A}_m(j\omega) \cdot W_m(j\omega) \approx \hat{A}_m^{ideal}(j\omega) \quad (15)$$

Fig. 4 shows the overall architecture of the proposed calibration technique. It can be seen that the proposed technique is not limited by the number of sub-ADC channels. If we assume that one of the channels is ideal, then only (M-1) channels need calibration.

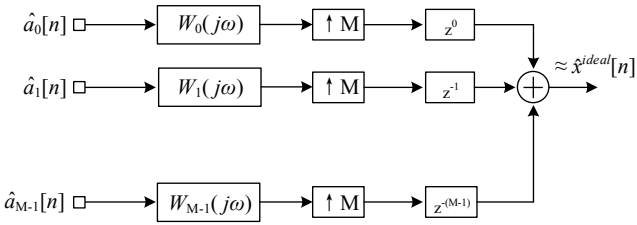

Fig. 4.  The proposed calibration architecture in M-channel TIADC.

## IV. POLYPHASE IMPLEMENTATION

Fig. 5 shows the polyphase implementation structure of the proposed technique. The key idea is to treat each sub-ADC as one L-channel TIADC. The data of each sub-ADC is broken up into L portions, after parallel calibration, the output data is reorganized into one stream, which is approximately equal to the ideal output stream. This way of processing can speed up the calculation rate by L times in the hardware calibration, achieving real-time calibration.

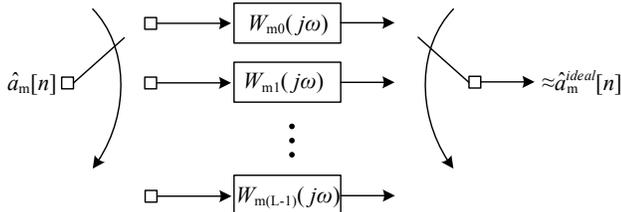

Fig. 5.  Polyphase implementation structure.

## V. EXPERIMENTAL RESULTS

### A. Simulation Results

To verify the efficiency of the proposed technique, simulations are carried out on a TIADC. Unless otherwise noted,

the channel number is two, signal word length is 12bits, the frequency of analog input signal is $0.019f_s$, the number of FIR taps is 30, the FIR coefficient word length is 30 bits, offset mismatches $\Delta o_m = 0$ (m=0, 1), gain mismatches and sample-time mismatches $\Delta g_m$ and $\Delta t_m$ (m=0, 1) are 0, 0.01 and 0, 0.01, respectively.

Fig. 6 shows output spectrum of the TIADC. The frequency spectra of mismatches are reduced after calibration, and the SINAD of signal is improved from 45 dB to 69 dB.

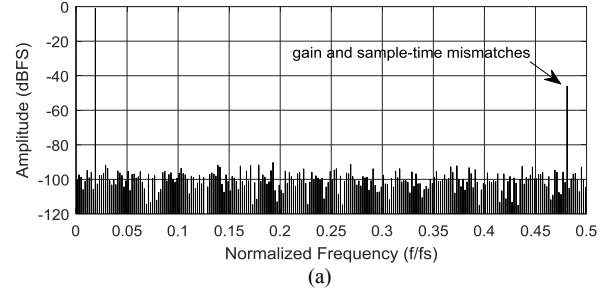

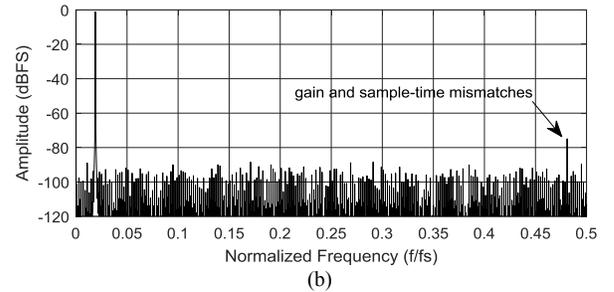

Fig. 6.  Single frequency signal spectrum in a two-channel TIADC simulation: (a) without and (b) with calibration.

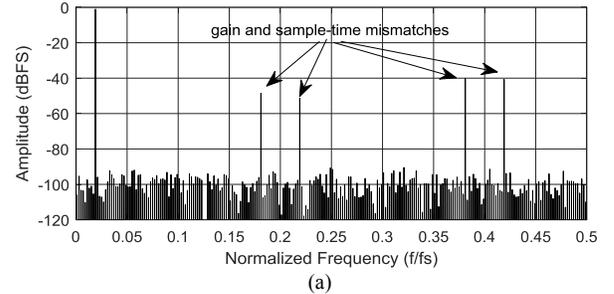

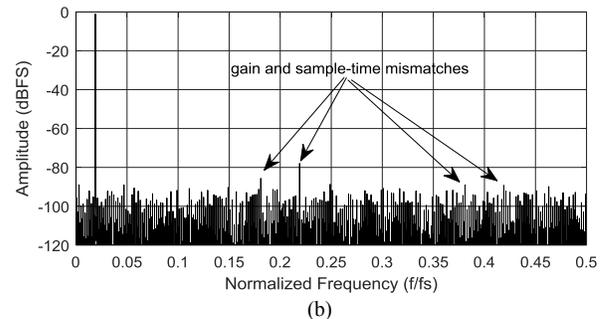

Fig. 7.  Single frequency signal spectrum in a five-channel TIADC simulation: (a) without and (b) with calibration.

Fig. 7 shows output spectrum of a five-channel TIADC. The offset mismatches $\Delta o_m = 0$ (m=0, 1, 2, 3, 4), gain mismatches and sample-time mismatches $\Delta g_m$ and $\Delta t_m$ (m=0, 1) are 0, 0.01, -0.01, 0.02, -0.02 and 0, 0.01, 0.02, -0.01, -0.02, respectively. The frequency spectra of mismatches are reduced after calibration, and the SINAD of signal is improved from 36 dB to 69 dB.

Fig. 8 shows output spectrum of the TIADC when input signal frequencies are $0.019f_s$, $0.133f_s$, $0.266f_s$, $0.399f_s$, respectively. The frequency spectra of mismatches are reduced after calibration, which proves the effectiveness of this technique in the wide-band range.

Fig. 9 shows SINAD versus filter coefficient word length. We can see that 24-bit FIR coefficient is sufficient to correct 12-bit signals in the case of slight mismatches. Fig. 10 shows SINAD versus the number of filter taps when sample-time mismatches $\Delta t_m$ (m=0, 1) are 0, 0.02.

Fig. 11 shows SINAD versus gain mismatch $\Delta g_1$ when sample-time mismatches $\Delta t_m$ (m=0, 1) are 0, 0 and the input signal frequency is $0.46f_s$. Fig. 12 shows SINAD versus sample-time mismatch $\Delta t_1$ when gain mismatches $\Delta g_m$ (m=0, 1) are 0, 0 and the input signal frequency is $0.19f_s$.

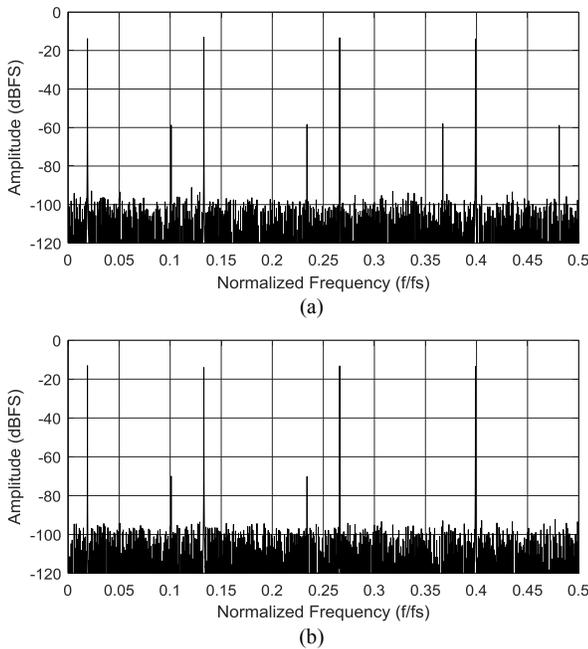

Fig. 8. Wideband signal spectrum in a two-channel TIADC simulation: (a) without and (b) with calibration.

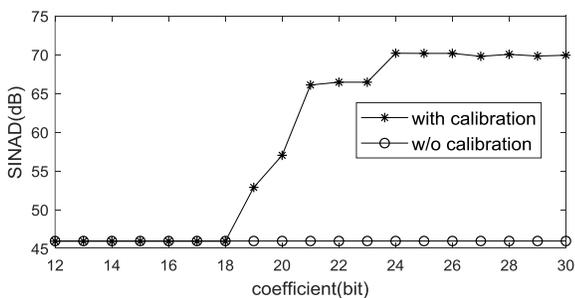

Fig. 9. SINAD versus filter coefficient word length with/without calibration.

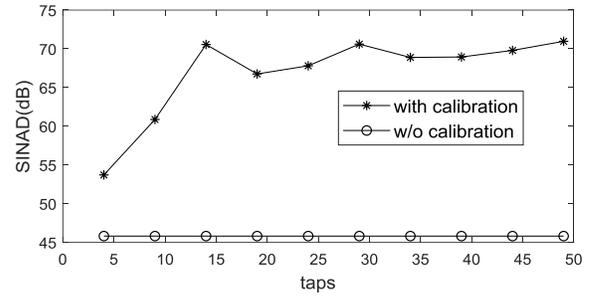

Fig. 10. SINAD versus filter taps with/without calibration.

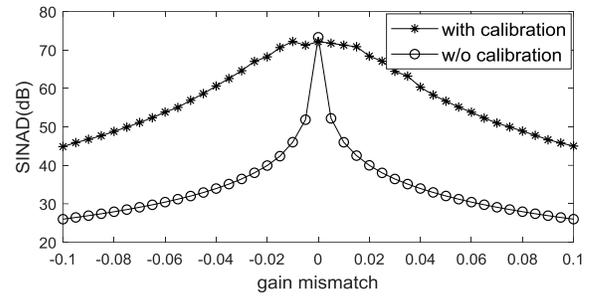

Fig. 11. SINAD versus gain mismatch with/without calibration.

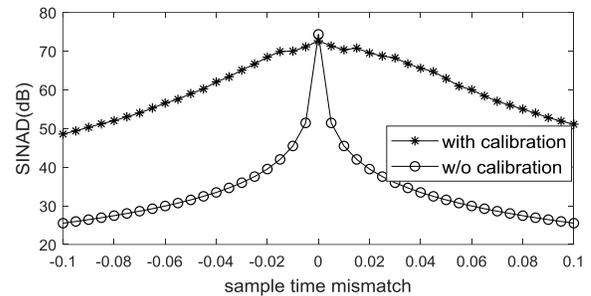

Fig. 12. SINAD versus sample-time mismatch with/without calibration.

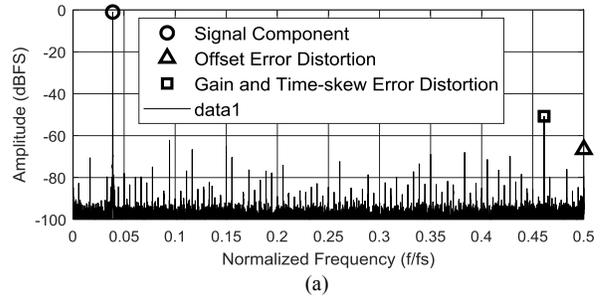

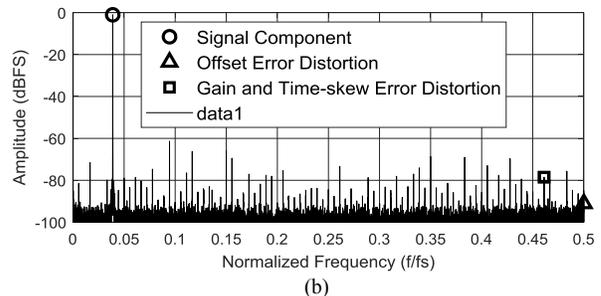

Fig. 13. A two-channel TIADC measured data output spectrum: (a) without and (b) with calibration.

*B. Measured Data Validation*

In addition, measured data from an actual two-channel TIADC, is carried out calibration. The $f_s$ of the TIADC is 1.8GSPS. The signal word length is 12bits, the frequency of analog input signal is $0.039f_s$, the number of FIR taps is 30, the FIR coefficient word length is 30 bits. Fig. 13(a) shows spectrum before calibration. After using the calibration technique, the spectrum is shown in Fig. 13(b). The frequency spectra of mismatches are reduced after calibration, and the SINAD of signal is improved from 47 dB to 53 dB.

## VI. CONCLUSION

This paper has described a new area-efficient all-digital technique for calibration of the mismatches in TIADCs. The proposed technique gives the first order approximation of the gain mismatches and sample-time mismatches, and employs first order approximation FIR filter banks to calibrate the sampled signal, which do not need large number of FIR taps. In the case of a two-channel 12-bit TIADC, the proposed technique improves SINAD of simulated data from 45dB to 69dB, and improves SINAD of measured data from 47dB to 53dB, while the number of FIR taps is only 30. In the case of slight mismatches, 24-bit FIR coefficient is sufficient to correct 12-bit signals, which makes it easy to implement this technique in hardware. In addition, this technique is not limited by the number of sub-ADC channels and can be calculated in parallel in hardware, these features enable this technique to be versatile and capable of real-time calibration.